\documentclass[8pt]{article}
\usepackage{arxiv}
\usepackage[utf8]{inputenc} 
\usepackage[T1]{fontenc}    
\usepackage{hyperref}       
\usepackage{url}            
\usepackage{booktabs}       
\usepackage{amsfonts}       
\usepackage{nicefrac}       
\usepackage{microtype}      
\usepackage{graphicx}		
\usepackage{amsmath}
\usepackage{amssymb}
\usepackage{widetext}
\usepackage[utf8]{inputenc}
\begin{document}
\title{Analysis of ring laser gyroscopes including laser dynamics}
\author{
 Angela D.V.~Di~Virgilio \\
  INFN Sez. di Pisa,\\
   Polo Fibonacci, Largo B Pontecorvo 3, I-56127 Pisa, Italy \\
\texttt{angela.divirgilio@pi.infn.it} \\
\And
Nicol\`o Beverini, Giorgio Carelli, Donatella Ciampini, Francesco Fuso, and Enrico Maccioni\\
  Universit\`a di Pisa,\\ Dipartimento di Fisica "E. Fermi", Largo B Pontecorvo 3, I-56127 Pisa, Italy
}

\maketitle







\maketitle
\begin{abstract}
Inertial sensors stimulate very large interest, not only for their application but also for fundamental physics tests. Ring laser gyros, which measure angular rotation rate, are certainly among the most sensitive inertial sensors, with excellent dynamic range and bandwidth.
Large area ring laser gyros are routinely able to measure fractions of prad/s, with high duty cycle and bandwidth, providing fast, direct and local measurement of relevant geodetic and geophysical signals. Improvements of a factor $10-100$ would open the windows for general relativity tests, as the GINGER project, an Earth based experiment aiming at the Lense-Thirring test at $1\%$ level. However, it is well known that the dynamics of the laser induces non-linearities, and those effects are more evident in small scale instruments. Sensitivity and accuracy improvements are always worthwhile, and in general there is demand for high sensitivity environmental study and development of inertial platforms, where small scale transportable instruments should be used. We discuss a novel technique to analyse the data, aiming at studying and removing those non-linearity. The analysis is applied to the two ring laser prototypes GP2 and GINGERINO, and angular rotation rate evaluated with the new and standard methods are compared. The improvement is evident, it shows that the back-scatter problem of the ring laser gyros is negligible with a proper analysis of the data, improving the performances of large scale ring laser gyros, but also indicating that small scale instruments with sensitivity of nrad/s are feasible.
\end{abstract}
\begin{twocolumn}

\section{Introduction}
Ring laser  gyroscopes (RLGs) are inertial sensors based on the Sagnac effect \cite{RSIUlli,CR,DIVIRGILIOCR}. They are largely utilised for inertial navigation, and applications in geodesy, geophysics and even for General Relativity tests are foreseen \cite{brz}. Since 2011 we are studying the feasibility of the Lense Thirring test at the level of $1\%$ with an array of large frame RLGs \cite{prd2011,angelo2017,angela2017}.
For that purpose it is necessary to push the relative accuracy of the Earth rotation rate measurement in the range from $1$ part in $10^9$ up to $1$ part in $10^{12}$.
RLG consists of a laser with a cavity comprising of three or four mirrors, depending if the cavity is triangular or square, rigidly attached to a frame; large  frame RLGs are utilised to measure the Earth rotation rate, being attached to the Earth crust. Because of the Sagnac effect, the two counter-propagating cavity modes have slightly different frequency, and the beat note of the two  beams is proportional to the angular rotation rate felt by the ring cavity. Large frame RLGs are the most sensitive instruments for inertial angular rotation measurements. The Sagnac frequency of a RLG is in fact proportional to the angular rotation rate $\vec{\Omega}$ which affects the apparatus:
\begin{eqnarray}
f_s =S \Omega \cos{\theta}\\
 S = 4\frac{A}{\lambda L} \nonumber
\label{uno}
\end{eqnarray}
where $A$ is the area of the ring cavity, $L$ is its perimeter, $\lambda$ the wavelength of the light, and $\theta$ is angle between the area versor of the ring and the orientation of $\vec{\Omega}$. For RLGs horizontally aligned (area versor vertical) $\theta$ is the colatitude angle, while for RLGs aligned at the maximum Sagnac signal $\theta = 0$. Eq. \ref{uno} connects $\Omega$ with  the scale factor $S$, which depends on the geometry, and $\lambda$, quantities than can be measured.
Further to sensitivity, other advantages of such instruments rely on the broad bandwidth,  which can span from kHz down to DC, and the very large dynamical range. In fact the same device can record microseismic events and high magnitude nearby earthquakes \cite{simonelli}, owing to the fact that the signal is based on the measurement of the beat note. It has been proven that large size RLGs with state of the art mirrors can reach the relative precision of $3$ parts in $10^9$ in one day integration time, for the Earth rotation rate measurement\cite{RSIUlli}.  If shot noise limited, the sensitivity scales with the second power of the size of the ring cavity.

The main limitation of RLG performances is given by the coupling between the two counter propagating laser modes. On each cavity mirror a little fraction of the two traveling waves is backscattered in the opposite direction. As a result of the interference between the reflected waves from each mirror, we have effective backscattering amplitude reflectivity $r_{1}$ and $r_{2}$  of the beam $1$ over the beam $2$ and of the beam $2$ over the beam $1$, respectively. We outline that the  interference of the back reflected waves, and consequently the values of  $r_{1,2}$ are very sensitive to any perturbation of the optical cavity geometry.
This coupling produces for small rotational rate a pulling of the Sagnac frequency  from the $f_s$ value given by eq. \ref{uno}, and eventually a locking of the two laser frequencies when the $f_s$ value become lower than $f_{lock}=r_{1,2}/ \pi$  \cite{stedman}. As a rough estimation, it is possible to evaluate  for a square cavity $f_{lock}= \frac{c\mu\lambda}{\pi d L}$, where $c$ is the velocity of light,  $d$ the diameter of the beam and $\mu$ the total scattered fractional amplitude at each reflection.

In order to ensure the functionality of small scale RLGs\cite{navigation}, mechanical dithering is usually implied  to increase the bias between the two modes and avoid locking. Large frame rings utilise the Earth rotation rate as bias. Any improvement in the accurate evaluation of the backscatter noise, and in general of the systematics induced by the non linear dynamics of the lasing process, is advantageous for increasing the performance of both large and small frame RLGs.
Presently there is large interest in this kind of device, large scale apparatus should further improve their sensitivity and accuracy for geodetic and fundamental physics application, and small scale and transportable devices at nrad/s sensitivity are required to improve the low frequency response of gravitational wave antennas, for the development of inertial platforms, and for seismology \cite{LIGO1, LIGO2, virgoAngela1, virgoAngela2, virgoAngela3}. 
 \\
The problem of the reconstruction of signals is a general one, and sophisticated filters can be developed to this aim. In the past we have addressed this problem utilizing Kalman filters, with whom we have obtained good results, but which were rather time consuming. At present, we have the necessity to analyze a very large set of data and to set up mathematical tools for the analysis with the aim not only to evaluate the sensitivity, but also to precisely identify specific issues in the setup which are limiting the sensitivity.
   \\
This paper presents a mathematical approach to measure the Sagnac frequency taking into account the laser dynamics.  This issue has been addressed several times \cite{stedman, hurst, beghi1, beghi2, bretenaker}, but no general solution exists. Analytical solutions can be derived in the case the backscattered light is equal in the two counter propagating modes, or the ratio between the intensities of the two modes is constant, conditions which are not fulfilled in the actual generation of RLGs \cite{beghi1}.
The discussion is composed of two main parts. The first one, after a short description of the RLG and the standard analysis approach,  describes the general RLG dynamics and reconstructs the Sagnac frequency taking into account the laser dynamics in the general case through a single analytical formula containing the laser  coefficients (Lamb coefficients), which can be separately evaluated based on experimental measurements. This formula can be further divided as linear sum of six contributions. One of the contributions, called $\omega_{s0}$,  is the dominant one, the others being small corrections. In the second part of the paper the implementation  of $\omega_{s0}$ is discussed and applied to data of the RLG prototypes GP2 and GINGERINO. The  other additional terms are not considered in the implementation, they will be subject of future work based on the data of GINGERINO. The appendix reports a short discussion about noise, and practical methods to identify portions of data which have to be discarded.

\section{Typical RLG and standard analysis method}
Fig. \ref{GeneralScheme} shows the typical lay-out of a square cavity RLG. The four mirrors are placed at the corner of the square ring, they are contained inside small vacuum chambers, which are connected by pipes.
\begin{figure}[htbp]
\centering
\includegraphics[width=\linewidth]{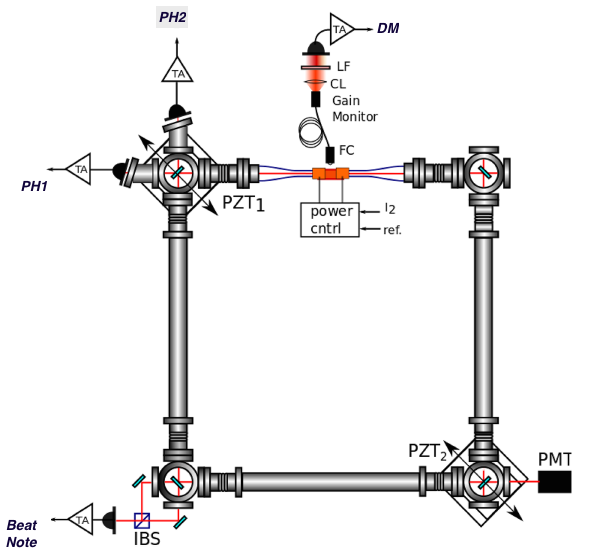}
\caption{Typical scheme of RLG with a square ring cavity.}
\label{GeneralScheme}
\end{figure}
The whole setup is vacuum tight and filled with Helium and an isotopic $50/50$ mixture of $^{20}$Ne and $^{22}$Ne. In one of the side the laser discharge is located to generate the plasma required for laser operation (top side). Piezoelectric translators are utilised to translate the mirrors, allowing a control of the RLG perimeter length. The Sagnac beat note signal is observed at one corner (bottom-left) by superimposing the two output beams on a photodiode. Other two photodiodes monitor at one of the output corners (top-left) the laser output power of the two beams ($PH_{1}$, $PH_{2}$). We will indicate them in the following as the \textit{mono-beam} signals. Another photodiode (\textit{discharge monitor}) records the fluorescence from the discharge, filtered around $633$ nm by a narrow width interferometer filter, which provides a rough indication of the density of excited atoms. In the usual operations, the plasma discharge is electronically controlled in order to keep constant the output power of one of the two mono-beam signal.
All these signals are acquired by an ADC with a frequency rate of a few kHz, suitable to well reconstruct them from DC up to the Sagnac frequency.

In the standard analysis, the Sagnac angular frequency $\omega_S$ is assumed to be equal to the instantaneous frequency $\omega_m$ , reconstructed from the interferogram by means of the Hilbert transform or of the standard AR2 recursive algorithms based on autocorrelation. The backscatter noise is usually subtracted by fitting the quantity $\frac{I_{S1} I_{S2}}{PH_1 PH_2} \cos{2\epsilon}$, where $\epsilon$ is the backscatter phase and $PH_{1,2}$ and  $I_{S1,S2}$ are the amplitudes of DC and $\omega_S$ spectral components of mono-beams 1 and 2 , respectively \cite{hurst}.

\section{Ring Laser dynamics, approximations and the stationary solution}
The present analysis is dedicated to the evaluation of the Sagnac signal taking into account the dynamics of the ring laser. The most general description of the RLG is based on the model developed by Aronowitz following the more general Lamb theory \cite{lamb,aronowitz}. The general equations are
\begin{widetext}
\begin{gather}
\dot{I}_1 = \frac{c}{L} ( \alpha _1 I_1 - \beta _1 I_1^2 - \theta _{12} I_1 I_2 +2 r_2 \sqrt{I_1 I_2} \cos (\psi +\epsilon ))
\label{generalm1}\\
\dot{I}_2 = \frac{c}{L} ( \alpha _2 I_2 - \beta _2 I_2^2 - \theta _{21} I_1 I_2 +2 r_1 \sqrt{I_1 I_2} \cos (\psi -\epsilon ))
\label{generalm2}\\
\begin{split}
\dot{\psi} = & \omega _s-\sigma _1+\sigma _2 -  \tau _{12} I_2+ \tau _{21}I_1 +\\
& - \frac{c }{L} (r_1 \sqrt{\frac{ I_1}{ I_2}} \sin (\psi -\epsilon )+r_2 \sqrt{\frac{ I_2}{ I_1}}  \sin (\psi +\epsilon ) )
\label{general}
\end{split}
\end{gather}
\end{widetext}
where $I_1$, $I_2$  are  the intra-cavity laser intensities expressed in the dimensionless "Lamb units"; $\psi$ and $\dot{\psi }$ are the instantaneous phase difference and its time derivative; index 1 and 2 refers  to the clockwise and counter-clockwise laser beam respectively. It is important to remind that all terms of  eq. \ref{generalm1},\ref{generalm2}, and \ref{general} are time depending. Here $\alpha_{1,2}$, $\sigma_{1,2}$, $\beta_{1,2}$, $\theta_{12,21}$, $\tau_{12,21}$ are the Lamb parameters. In particular, $\alpha_{1,2}$, $\sigma_{1,2}$ are the threshold gain minus losses, $\beta_{1,2}$ is the self saturation, $\theta_{12,21}$, $\tau_{12,21}$ describe cross-(mutual-)saturation.  The Lamb theory involves a large number of parameters, however,  the special mixture of two isotopes of Neon and the working point close to the laser threshold allow adoption of a simplified model \cite{beghi1,beghi2,harmonic}. In our present analysis, we assume $\beta_1 = \beta_2 = \beta$, and $\theta_{21} = \theta_{12}=\theta$.  This assumption is justified by the fact  that our RLGs operate close to threshold in mono-mode regime (for operation near multi-mode regime, a further approximation is feasible). In the following $\theta$ will be neglected, owing to the mono-mode operation. \\
Without loss of generality we can define $\delta_{ns} = \sigma _2 - \sigma _1 + \tau _{21} I_2- \tau _{12}I_1 $, which is usually referred to as null shift; it is generally accepted that $\delta_{ns}$ is a small quantity to be neglected \cite{stedman,beghi1,lamb}. In the present analysis it will not be neglected: $\delta_{ns}$ will be considered a perturbation of $\dot{\psi}$, defining a new variable  $\dot{\psi_0} \simeq \dot{\psi}-\delta_{ns}$ ( $\dot{\psi}$ being the frequency effectively measured by the interferogram, called also $\omega_m$).\\
Assuming that RLG is at the steady state \cite{beghi1},  the solutions are the following:
\begin{gather}
\begin{split}
I_1( t)\simeq & \frac{\alpha _1}{\beta }+\frac{2 \sqrt{\alpha _1 \alpha _2} r_2 (\frac{L \omega _s \sin (t \omega _s+\epsilon )}{c}+\alpha _1 \cos (t \omega _s+\epsilon ))}{\beta  (\alpha _1^2+\frac{L^2 \omega _s^2}{c^2})}+\\
& -\frac{2 c r_1 r_2 \sin (2 \epsilon )}{\beta  L \omega _s}\\
I_2 (t)\simeq & \frac{\alpha _2}{\beta }+\frac{2 \sqrt{\alpha _1 \alpha _2} r_1 (\alpha _2 \cos (\epsilon -t \omega _s)-\frac{L \omega _s \sin (\epsilon -t \omega _s)}{c})}{\beta  (\alpha _1^2+\frac{L^2 \omega _s^2}{c^2})}+\\
& +\frac{2 c r_2 r_1 \sin (2 \epsilon )}{\beta  L \omega _s}\\
\psi_0 (t)\simeq & \frac{c (\sqrt{\frac{\alpha _1}{\alpha _2}} r_1 \cos (\epsilon -t \omega _s)+\sqrt{\frac{\alpha _2}{\alpha _1}} r_2 \cos (t \omega _s+\epsilon ))}{L \omega _s}+\\
& +t (\omega _s-\frac{2 r_1 r_2 (\frac{c}{L})^2 \cos (2 \epsilon )}{\omega _s})
\end{split}
\end{gather}
The validity of the above solutions  has been previously tested with a Monte Carlo simulation and with the experimental data of the RLG G-Pisa, which was a $5.40$ m perimeter RLG \cite{beghi2,Kalman61}. Here the validity of previous results  is taken for granted and  $\omega_s$  is analytically expressed. \\
Assuming that the parameters are constant in the time interval between  $t$ and $t + \delta t$, we have $\psi_0(t+\delta t)-\psi_0(t) = \omega_m \delta t - \delta_{ns} \delta t$.
 From the above relation it is straightforward to deduce that,  at the first order in $\delta t$:
\begin{equation}
(\omega _m -\delta_{ns}) \delta t  \simeq (\omega _s + \frac{K(t)}{L} -\frac{2 c^2 r_1 r_2 \cos (2 \epsilon )}{L^2 \omega _s})\delta t .
 \label{fase}
\end{equation}
In the above equation  $\omega_s$ is the Sagnac angular frequency, the quantity we are looking for, and we have conveniently defined $K(t)$:
\begin{equation}
K(t) = \sqrt{\frac{\alpha _1}{\alpha _2}} c r_1 \sin (\epsilon -t \omega _s)-\sqrt{\frac{\alpha _2}{\alpha _1}} c r_2 \sin (t \omega _s+\epsilon ),
\end{equation}
$K(t)$ contains  oscillatory terms at the Sagnac frequency $\omega_s$.
Considering that $\omega_s$ is almost constant, for frequency much below the Sagnac frequency, it is possible to  look for approximated solutions.
Eq. \ref{fase}  can be written as:
 \begin{equation}
\omega_s = \frac{\omega_m}{2}  + \sqrt{\frac{8 c^2 r_1 r_2 \cos (2 \epsilon )+(K-L (\omega_m +\delta_{ns}))^2}{4 L^2}}-\frac{K}{2L}+\frac{\delta_{ns}}{2}
\label{fase1}
\end{equation}
where we dropped the time dependence in $K$. The occurrence of the oscillations of $K$ at the Sagnac frequency makes the evaluation of $\omega_s$ non trivial. The average value of $K$ is very small for frequencies much below Sagnac frequency,  since the average value of sinus and co-sinus oscillating at the Sagnac frequency goes to zero for frequency much below $f_s$.
$\omega_s$ can be found with eq.\ref{fase1}, provided that  $r_1$, $r_2$, $\epsilon$, $\delta_{ns}$, and the average value of $K$ are available, which is in principle feasible utilising the mono-beam signals and the measured losses of the cavity  and employing numerical recursive methods to evaluate $K$.  \\

In the following eq. \ref{fase1} will be decomposed in several pieces, which can be separately evaluated.
In any case, when  $| K | \ll L \omega_m$ and  $ \delta_{ns} \ll  \omega_m$, eq. \ref{fase1}
can be expanded  in $K$ and $\delta_{ns}$ at first and second order, obtaining:
\begin{align}
\omega_s & \simeq  \omega_{s0} +\omega_{ns1} +\omega_{ns2}  + \omega_{K1}+\omega_{K2}+\omega_{nsK} 
\label{general1}\\
\omega_{s0}& =  (\frac{1}{2} \sqrt{\frac{8 c^2 r_1 r_2 \cos (2 \epsilon )}{L^2}+\omega_m ^2}+\frac{\omega_m }{2} )
\label{general0}\\
\omega_{ns1} &= -\delta_{ns}\times(\frac{\omega_m }{2 \sqrt{\frac{8 c^2 r_1 r_2 \cos (2 \epsilon )}{L^2}+\omega_m ^2}}+\frac{1}{2})\nonumber\\
\omega_{ns2} &= \delta_{ns}^2\times\frac{2 c^2  r_1 r_2 \cos (2 \epsilon )}{(8 c^2 r_1 r_2 \cos (2 \epsilon )+L^2 \omega ^2) \sqrt{\frac{8 c^2 r_1 r_2 \cos (2 \epsilon )}{L^2}+\omega_m ^2}}\nonumber\\
\omega_{K1} &=  K\times (-\frac{\omega_m }{2 L \sqrt{\frac{8 c^2 r_1 r_2 \cos (2 \epsilon )}{L^2}+\omega_m ^2}}-\frac{1}{2 L} )\nonumber \\
\omega_{K2} &=  K^2\times\frac{2 c^2 r_1 r_2 \cos (2 \epsilon ) \sqrt{\frac{8 c^2 r_1 r_2 \cos (2 \epsilon )}{L^2}+\omega_m ^2}}{ (8 c^2 r_1 r_2 \cos (2 \epsilon )+L^2 \omega_m ^2 ){}^2}\nonumber\\
\omega_{ns K}&= \frac{ \delta_{ns} K}{2\sqrt{8 c^2 r_1 r_2 \cos{2 \epsilon}+L^2{\omega_m}^2}}\nonumber
\end{align}
Eq. \ref{general1} is composed of $6$ terms, which  can be independently evaluated, and analysed. Careful evaluation is necessary for $\omega_{ns1,2}$, $\omega_{K1,2}$, since the determination of the parameters $\beta$, $\sigma_1$, $\sigma_2$, $\tau_{12}$, and $\tau_{21}$, which are function of the beam area $a$, the output power, the mirrors transmission and the total losses $\mu$, is required. The mathematical relationships to evaluate those terms can be found in previous papers \cite{hurst,beghi1,beghi2}.\\
The reconstruction of those terms  will be addressed in future work, and applied to  the analysis of the data of GINGERINO.

In the following the implementation of the first term $\omega_{s0}$  will be specialised for data acquired with large frame RLGs and compared with the standard analysis method. Backscatter noise is accounted for, and it has been checked that the standard method to subtract the backscatter noise can be derived from eq. \ref{general0} assuming $\frac{8 c^2 r_1 r_2 \cos (2 \epsilon )}{L^2} \ll \omega_m^2$ and expanding at first order.

\subsection{Application to the actual data}
The analysis described in the following will take into account data streams at normal operation and far from transients of the laser as mode jumps and split modes. Appendix A describes methods to identify and eliminate those portions of data.
As already said eq. \ref{general1} is valid for $K \ll L \omega_m$; referring to our smaller prototype G-Pisa (perimeter $5.40$ m), and utilising published parameters \cite{beghi1}, we obtain $K$ $\sim 6$ rad m/s, to be compared with $\omega_m L\sim 3600$ rad m/s: consequently eq. \ref{general1} is valid.
We underline that quoted values are conservative ones, since $K$ depends on the mirror quality and the size of the ring. The prototype G-Pisa was smaller than GP2 and GINGERINO and equipped with less performing mirrors. Determining $\omega_{s0}$ requires in turn to evaluate $r_1$ and $r_2$.
 Following previous works \cite{beghi1,beghi2}, it is possible to link such quantities with available measured data:
\begin{eqnarray}
r_1=\frac{  I_{S2} \omega _m}{\frac{ 2 c \sqrt{ PH_{1}  PH_{2}}}{L}}\\
r_2=\frac{  I_{S1} \omega _m}{\frac{ 2 c \sqrt{ PH_{1}  PH_{2}}}{L}}
\end{eqnarray}
 with all symbols already defined. Similarly, the relative phase $\epsilon$ is found comparing the mono-beams
 signal at the Sagnac frequency.
 All above quantities are commonly used in the standard analysis \cite{hurst,harmonic,90day}.
Substituting and simplifying, it is straightforward to show that:
\begin{equation}
\omega_{s0} = \frac{1}{2} \sqrt{(1+\xi) \frac{ 2  I_{S1}   I_{S2} \omega _m^2 \cos (2 \epsilon )}{ PH_{1}  PH_{2}}+\omega _m^2}+\frac{\omega _m}{2}
\label{approx}
\end{equation}
The term $\xi$ ($\xi \ll 1$) has been added in order to take into account inaccuracies on the mono-beams signals.  It is important to remark that the quantities $PH_1$, $PH_2$, and $I_{S1}$ and $I_{S2}$ refer
to the laser power inside the optical cavity, while measured ones are obtained utilising the power transmitted outside the cavity. Since Eq. \ref{approx} exploits the ratios, in principle it is not  affected by the measurement scheme, and the voltage output of the photodiodes can be used.\\
However, it is necessary to consider the presence of noise in the mono-beams signals, which can be due to the inherent noise of the photodiodes or by the discharge fluorescence, which cannot be completely removed. The related noise affects the evaluation of $\omega_{s0}$ done with eq. \ref{approx}. Therefore, in order to  have the possibility to correct it with common statistical methods, the term $\xi$ has been added. Expanding at first order in $\xi$, we obtain:
\begin{eqnarray}
\omega_{s0} = \frac{1}{2} \sqrt{\frac{ 2 \omega_m^2 I_{S1}   I_{S2} \cos (2 \epsilon )}{  I_{1}  I_{2}}+\omega _m^2}+\frac{\omega _m}{2} + \omega_{s \xi}\\
\label{formula}
\omega_{s \xi} = \xi\times\frac{  I_{S1}   I_{S2} \omega_m ^2 \cos (2 \epsilon )}{2  I_{1}  I_{2} \sqrt{\frac{ 2  I_{S1}   I_{S2} \omega _m^2 \cos (2 \epsilon )}{ I_{1}  I_{2}}+\omega _m^2}}
\label{xi}
\end{eqnarray}
It is straightforward to evaluate the term $\omega_{s0}$, while the corrective one,
$\omega_{s\xi}$, must be evaluated by fitting the parameter $\xi$. Remarkably, the above relation does not contain any Lamb parameter of the laser and can therefore be determined without knowledge of such parameters.

\section{Reconstruction of $\omega$\textsubscript{s0} for GP2 and GINGERINO}
Data acquired by our prototypes GINGERINO and GP2 are utilised. GP2 is an apparatus with comparatively low quality mirrors and located in a noisy environment \cite{belfi0, GP2new}, while GINGERINO is located in a very quiet place \cite{90day,RSI2016}, and is presently equipped with state of the art mirrors.
We remark that GINGERINO is free running, the geometry is not controlled and long time operation and high duty cycle ($>90\%$) are possible since it is located in the underground Gran Sasso laboratory, which exhibits high natural thermal stability (typically $\sim 0.01$ $^o$C in one day).

\subsection{Comparison of standard and new analysis}
Fig. \ref{OldNew} shows the comparison between the Sagnac frequency from GINGERINO data reconstructed with the standard method (referred to as $\omega_m$) and the one presented here. It is interesting to observe that the average value of the frequency $\omega_{s0}$ is higher, this is what we expect when the noise is dominated by backscatter. In such conditions, frequency is shifted upward ('pull') \cite{stedman}.
The average values are different for the two analysis methods; as far as GINGERINO is concerned, the difference is quite small, for example the analysis of $24$ days in November $2018$ gives a relative difference of $6\times10^{-5}$, with $<\omega_{s0} >$ evaluated by the method presented here a bit larger than  $<\omega_m>$. Since the absolute orientation of the RLG is unknown, in both cases the measured Sagnac frequency is compatible with the expected one assuming an inclination of  $\sim 1.5$ mrad northwards with respect to the horizontal plane.\\
\begin{figure}[htbp]
\centering
\includegraphics[width=\linewidth]{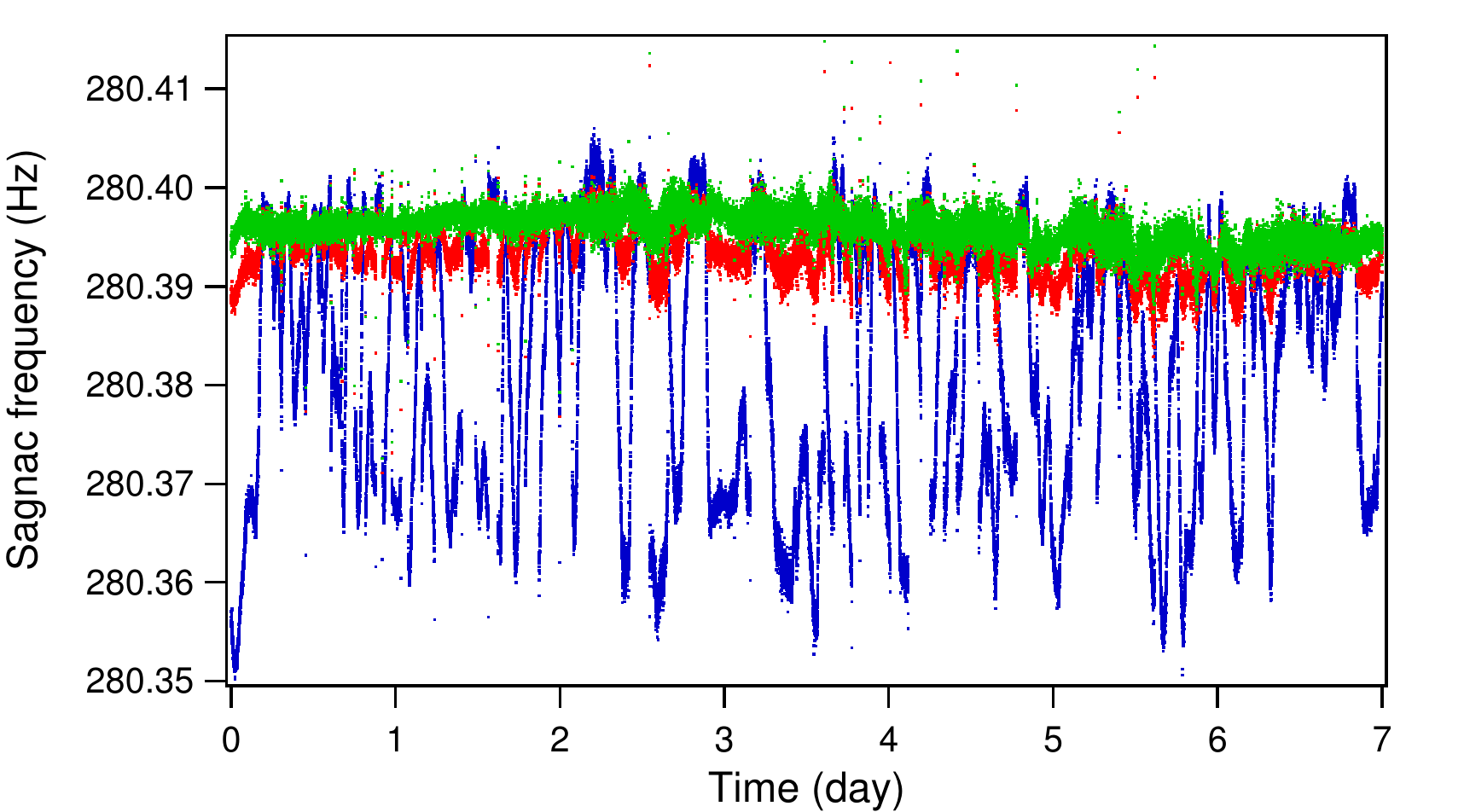}
\caption{Comparison of the old and new analysis of GINGERINO data. Blue trace: standard method with Hilbert transform; red trace: data corrected using eq. \ref{formula}; green trace: data corrected after fitting for parameter  $\xi$ (eq. \ref{xi},
in the fit $\xi=0.16$).}
\label{OldNew}
\end{figure}
Large frame RLGs are instruments dedicated to the study of phenomena with typical frequency below $20$ Hz;
we have checked that  the two methods are equivalent in the high frequency band of interest. Fig. \ref{OldNew20Hz}, showing the power spectral density (PSD) as a function of frequency, demonstrates that, for GINGERINO and above $200$ mHz, the difference between the two methods is less than $0.1$nrad/s in 1 second measurement. This comparison shows that the new analysis is not introducing extra noise above $200$ mHz at this level of sensitivity. It has been also checked that the old method, which  estimates and subtracts the backscattering effect through a linear fitting procedure, provides results distributed with width similar to $\omega_{s0}$, and, as already said, slightly different mean value. We outline that systematics of the laser dynamics include non linear terms, which in principle cannot be eliminated with  linear methods. Then, the standard method, being linear, cannot guarantee a full correction. \\
\begin{figure}[htbp]
\centering
\includegraphics[width=\linewidth]{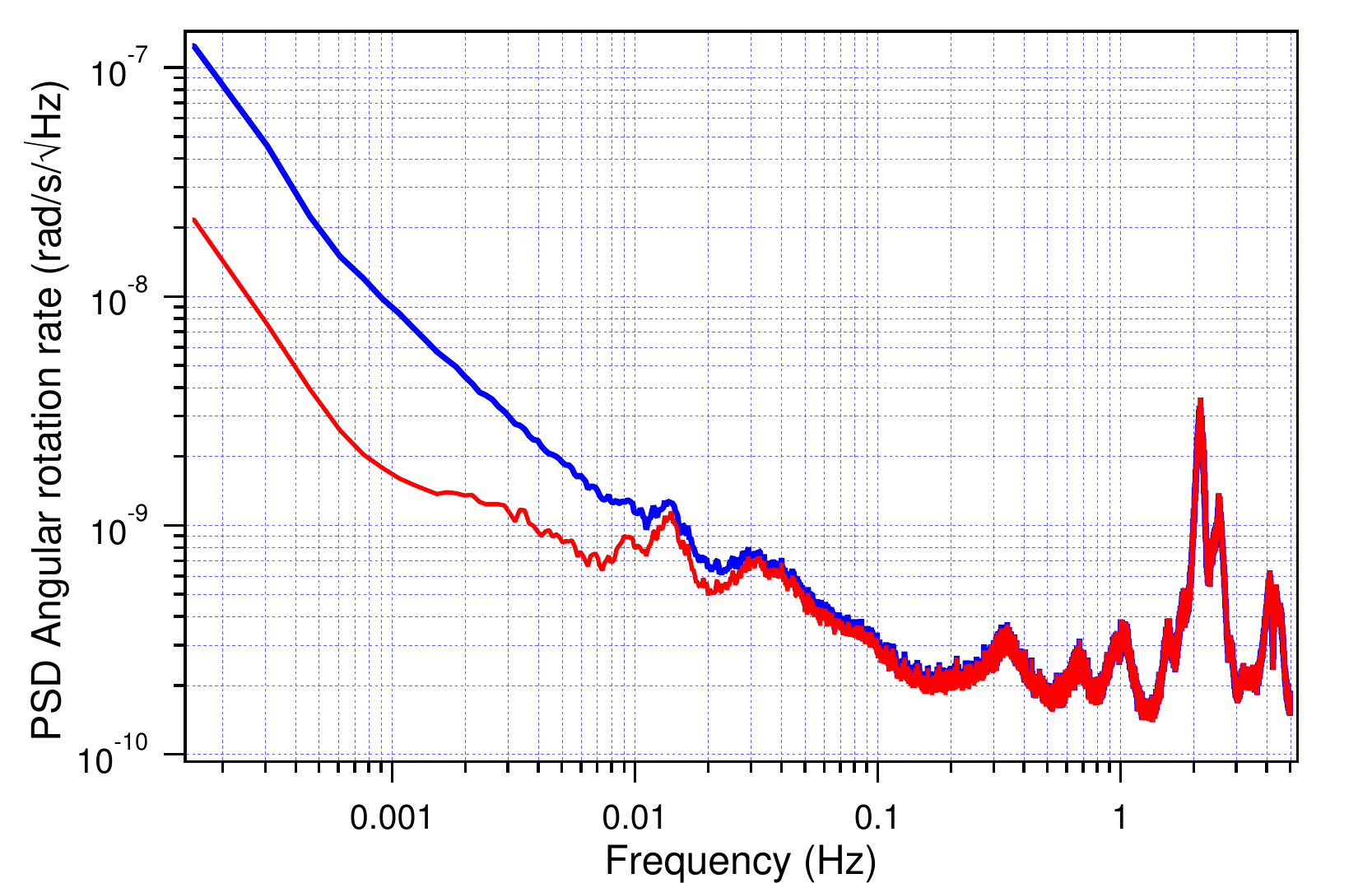}
\caption{Comparison of the power spectral density utilising the standard (blue data) and new analysis (red data). The two PSD are approximately equal above $200$mHz, the main differences are at low frequency.}
\label{OldNew20Hz}
\end{figure}
The systematics of RLG depends on the size and the mirror quality, large frame RLGs are usually closer to behave in an ideal manner. For reduced size RLG and when the mirrors are not top quality, deviations from the ideal case are  more relevant. This is the case of our prototype GP2.
Fig. \ref{histo} shows the histogram of the Sagnac frequency data of GP2 analysed with the two methods.
The standard analysis leads to a broader distribution and the mean value is compatible with a mean rotation frequency  $1$ Hz higher than expected.
GP2 is oriented at the maximum signal, so its response should be close to (and never higher than) the Earth rotation rate.
The new method gives an averaged rotation rate  $\Omega = 7.2922\times10^{-5}$ rad/s, in agreement with the Earth rotation rate $\Omega_\oplus =  7.292115\times 10^{-5}$ rad/s.
With the new analysis the average rotation rate is evaluated with a relative systematic error of $1$ part in $10^{-4}$, while with $6$ part in $10^{-3}$ with the standard analysis: a factor $60$ improvement in accuracy has been achieved. The present result is very similar to the one obtained in previous analysis based on the Kalman filter in term of accuracy and sensitivity.
In both cases the best sensitivity was of the order of a few nrad/s with tens of seconds integration time \cite{beghi2}.
\begin{figure}[htbp]
\centering
\includegraphics[width=\linewidth]{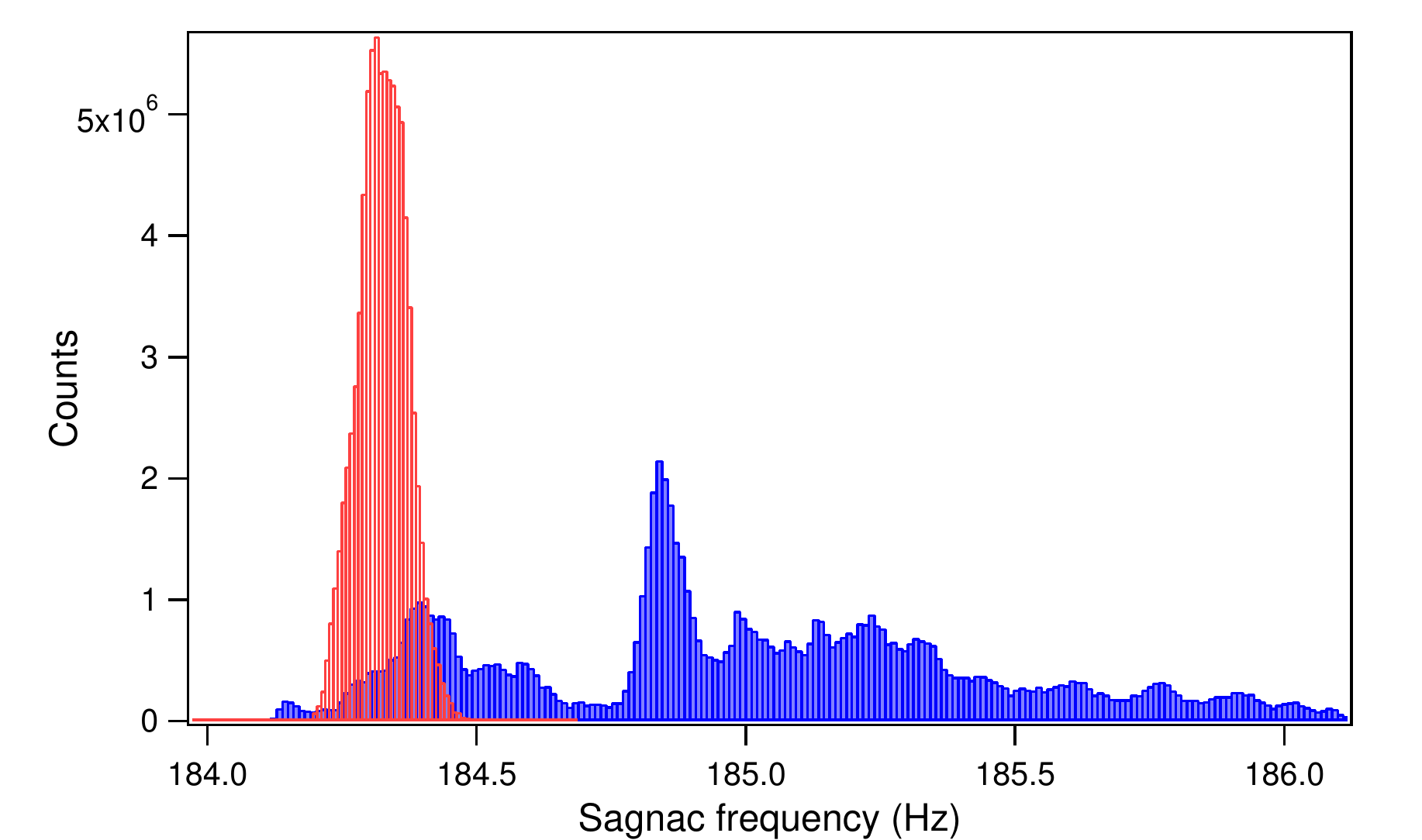}
\caption{Comparison of the histograms of the Sagnac frequency estimated with the standard method (blue) and by the new one (red). Clearly the new method leads to a narrower and more Gaussian-like distribution, with  mean  value $184.29$  Hz.}
\label{histo}
\end{figure}

It is puzzling to note that with the standard analysis method GP2 is showing higher than expected Sagnac frequency. A possible explanation is given within our mathematical approach to measure the Sagnac frequency. GP2 has been designed to test the geometry control developed for GINGER and based on diagonal length measurements; data with and without geometry control have been compared and it has been checked that, with adequate analysis, sensitivities are comparable \cite{GP2new}.
Fig. \ref{ConPh} compares the fringe contrast (TOP) and the  relative phase $\epsilon$ (BOTTOM) for GP2 data taken during the geometry control test. This was achieved by keeping constant the length of the two diagonals within 80 nm \cite{GP2new}. Fig. \ref{ConPh} shows that mode jumps occur in order to keep $\epsilon$ close to $\pm \frac{\pi}{2}$, this in support of the fact that GP2 has quite large backscatter light, and stable operation is favourite when $\epsilon$ is such that the two modes have an extra shift of $1$ Hz, also called "dissipative coupling regime" \cite{stedman}.
It is not perfectly clear from the theory why this regime takes place, but it is a matter of fact that the coupling between the two modes decreases increasing the bias frequency, accordingly stable operation is favourite. Another consequence is that mode jumps occur also to keep the relative phase close to a certain range, not only to compensate changes of the perimeter. It has been checked that GINGERINO exhibits all values of $\epsilon$.
\begin{figure}[htbp]
\centering
\includegraphics[width=\linewidth]{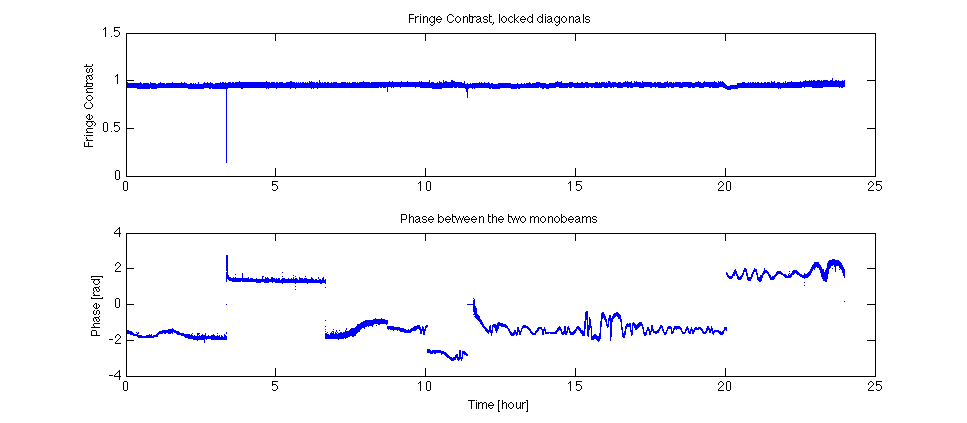}
\caption{TOP: Fringe contrast during geometry control of GP2, a few features suggesting mode jumps are shown. BOTTOM: the relative phase between the two modes $\epsilon$ is shown. In correspondence of the jumps there is a rapid change in phase. }
\label{ConPh}
\end{figure}
 \section{Conclusion}
 Systematics induced by the non linear dynamics of the laser, mainly due to backscatter light, induces non linear terms in the output of high sensitivity RLGs,  severely limiting the development of RLGs with sensitivity of the order of nrad/s level, which in principle should have a large range of applications.
An analytical method, suitable to reconstruct the Sagnac frequency $\omega_{s}$ taking into account the laser dynamics, has been developed in the general case in which the two backscattered beams are not  equal, and the ratio between the power of the two counter-propagating modes is not constant. The application of this formula requires the knowledge of the fractions of backscattered waves, and the laser parameters $\alpha_{1,2}$, $\sigma_{1,2}$, $\tau$ and $\beta$, all quantities which can be evaluated.  In the present theory the term $\theta$ is not considered, this term takes into account the multimode operation, and can be neglected in the description of high sensitivity RLGs which operates mono-mode close to threshold.\\
Expanding in series at first and second order it is possible to divide the general formula as the sum of six terms which can be separately evaluated. The analytical expansion for the whole set of $6$ terms is reported.  \\
The term called $\omega_{s0}$, which is the dominant one and does not contain any laser parameter, is evaluated in details and expressed as a function of the available measurements; this term has been evaluated for the two RLG prototypes GINGERINO and GP2, and compared with the standard analysis method.  The advantage of the new approach is evident: not only the width of the distribution is reduced, but the reconstructed Sagnac frequency is more accurate and in better agreement with the expected value. In short, $\omega_{s0}$ eliminates the so called backscatter noise, which is the dominant systematics especially for small and medium size RLG.\\
The GP2 prototype has more backscatter light,  because it is smaller and has lower quality mirrors with respect to GINGERINO. In this case the standard method evaluates the Earth rotation rate with a relative systematic error of $6$ part in $10^{-3}$, while in the new way $1$ part in $10^{-4}$ is obtained, a factor $60$ improvement in accuracy, with a sensitivity in the range of $2$ nrad/s with tens of seconds integration time.
This work opens up the window for the development of high sensitivity transportable RLGs, for applications in seismology, as environmental monitors  to improve the low frequency performance of the test mass suspension  of the gravitational wave antennas, and for the development of inertial platforms in general. Further efforts will be devoted to analyse the data of GINGERINO using the full set of terms.

 \appendix
 \section{Discussion about the noise}
Since we deal with high sensitivity measurements, it is important to estimate the noise injected in the evaluation of $\omega_{s0}$. $PH_i$ and $I_{Si}$ ($i,1,2$) are utilised to evaluate $\omega_{s0}$, and their noise will contribute to the total noise budget. In general all measurements of these quantities are limited by shot noise of the power collected by the photodiode $\delta_i, i=1,2$, and the total noise is the incoherent sum of the photodiode noise.
The contribution of each term, $\delta PH_1$ and $\delta I_{S1}$ gives:
\begin{eqnarray}
\delta I_{S1} \sim \frac{\delta_1   I_{S2} \omega ^2 \cos (2 \epsilon ))}{8  PH_{1}^2 PH_{2} \sqrt{\frac{I_{S1} I_{S2} \omega ^2 \cos (2 \epsilon )}{2 PH_{1} PH_{2}}+\omega ^2} }\\
\delta PH_{1} \sim \frac{\delta_{1} I_{S1} I_{S2}  \omega ^2 \cos (2 \epsilon )}{8 {PH_{1}}^2 PH_{2} \sqrt{\frac{I_{S1} I_{S2} \omega ^2 \cos (2 \epsilon )}{2 {PH_{1}} PH_{2}}+\omega ^2}}
\end{eqnarray}
 ($I_2$ and $I_{S2}$ are obtained changing 1 with 2 in a symmetric way). \\
Usually for top quality mirrors losses are minimised, but there are no requirements for the transmission. In order to minimise the contribution of the mono-beams to the total noise the optimal choice would be to have  transmission of the same order of the losses, at least for one of the mirrors (one output only is enough since in order to evaluate $\epsilon$ it is necessary to observe the two mono-beams transmitted by the same mirror). Care is also necessary in order to avoid small spurious reflections from one of the windows of the vacuum chamber, and narrow band filters are necessary in order to reduce the spurious signal from the discharge fluorescence. In any case, especially the measurement of the two terms $PH_{1,2}$ could be a real limitation for the very low frequency measurements, since they are DC quantities affected by the well known $1/f$ noise of any electronic device.\\

\section{The fringe contrast: a suitable tool to remove bad portions of data}

In large frame RLG attached to the Earth crust the Sagnac frequency is usually  above $100$ Hz,
and  it is determined by the Earth rotation rate, which is almost constant in time. The relative phase $\epsilon$ is slowly changing, since the cavity is rigid. In general, unless the geometry is electronically controlled, it happens that the RLG changes its operational points; accordingly the wavelength changes separately for both modes, or for one only, and mode jumps or split mode operations occur.
Split mode operation occurs from time to time; in principle the split mode regime provides good measurement of $\omega_s$, but in this case $\omega_m = \omega_s + 2\pi FSR$ (FSR, Free Spectral Range), and data acquisition at high rate and accurate knowledge of the perimeter are necessary. In the present analysis data affected by split mode operation have been disregarded. Mode jumps are very fast transients, affecting only few seconds of data acquisition. During these discontinuities the RLG is not at the stationary condition, so portions of data have to be discarded. The observation of the fringe contrast provides a very efficient tool to identify and eliminate those imperfections.
Fig. \ref{GINGERINOFC} shows corresponding split mode and mode jumps. Sometime some instabilities in the operation are visible before the  mode jump takes place. Fig. \ref{TypMJ} shows the typical behaviour of the mode jump.
\begin{figure}[htbp]
\centering
\includegraphics[width=\linewidth]{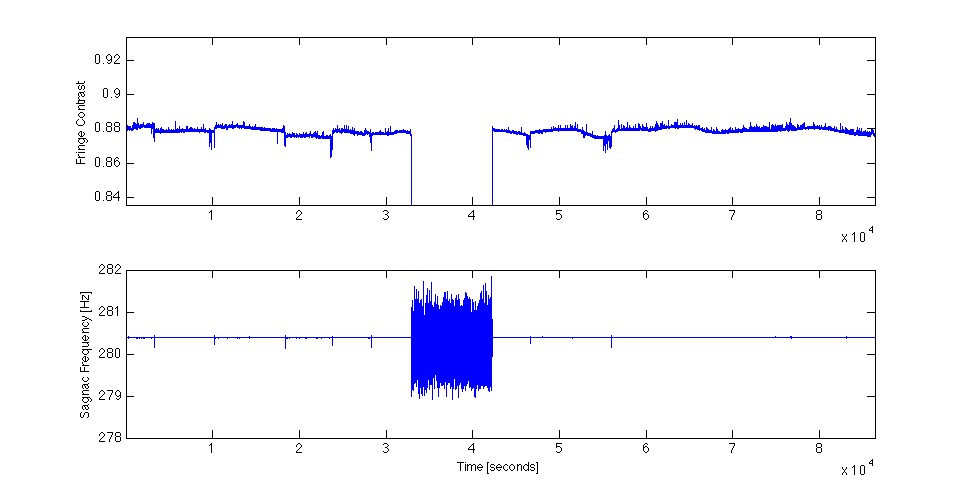}
\caption{TOP: typical fringe contrast, the mode jumps are evident, it is also clear that instabilities occur before the mode jumps, in the middle there is a split mode operation of the duration of 2.6 hours. BOTTOM: the corresponding Sagnac frequency.}
\label{GINGERINOFC}
\end{figure}
\begin{figure}[htbp]
\centering
\includegraphics[width=\linewidth]{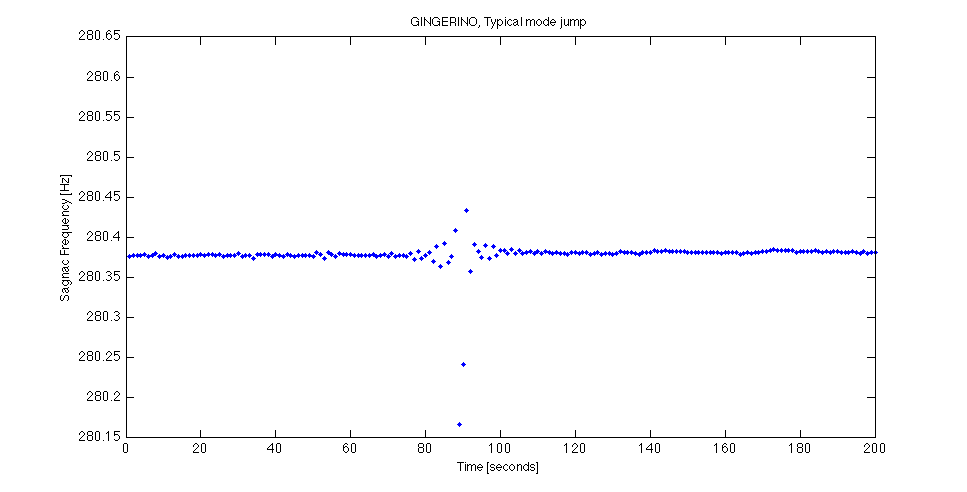}
\caption{GINGERINO Sagnac frequency around a typical mode jump. }
\label{TypMJ}
\end{figure}
\end{twocolumn}
\end{document}